\documentclass[conference]{IEEEtran}
\IEEEoverridecommandlockouts
\usepackage{cite}
\usepackage[font=small]{caption}
\usepackage{caption}
\usepackage{subcaption}
\usepackage[ruled,linesnumbered]{algorithm2e}
\usepackage[ruled,linesnumbered]{algorithm2e}
\usepackage{algpseudocode}
\def\BibTeX{{\rm B\kern-.05em{\sc i\kern-.025em b}\kern-.08em T\kern-.1667em\lower.7ex\hbox{E}\kern-.125emX}}

\makeatletter
\def\BState{\State\hskip-\ALG@thistlm}
\makeatother

\usepackage{array}
\newcolumntype{P}[1]{>{\centering\arraybackslash}p{#1}}
\usepackage{mathtools}
\usepackage{multirow}

\usepackage{graphicx}
\usepackage{dsfont}
\usepackage{amsmath}
\usepackage[hidelinks]{hyperref}
\usepackage{amssymb}
\usepackage{amsthm}
\usepackage{color}
\newcommand\numberthis{\addtocounter{equation}{1}\tag{\theequation}}

\newtheorem*{proof*}{\hspace{0.8cm}{\it{Proof}}}

\DeclareMathOperator*{\argmin}{argmin}
\ifCLASSINFOpdf

\else

\fi
\begin{document}
	\title{Federated Learning for Cellular-connected UAVs: Radio Mapping and Path Planning}
	\author{Behzad~Khamidehi and Elvino S. Sousa
		\\
		Department of electrical and computer engineering, university of Toronto, Canada.\\ 
		Emails: {b.khamidehi@utoronto.ca, and es.sousa@utoronto.ca} 
		}
	\maketitle

	\begin{abstract}

To prolong the lifetime of the unmanned aerial vehicles (UAVs), the UAVs need to fulfill their missions in the shortest possible time. In addition to this requirement, in many applications, the UAVs require a reliable internet connection during their flights. In this paper, we minimize the travel time of the UAVs, ensuring that a probabilistic connectivity constraint is satisfied. To solve this problem, we need a global model of the outage probability in the environment. Since the UAVs have different missions and fly over different areas, their collected data carry local information on the network's connectivity. As a result, the UAVs can not rely on their own experiences to build the global model. This issue affects the path planning of the UAVs. To address this concern, we utilize a two-step approach. In the first step, by using Federated Learning (FL), the UAVs collaboratively build a global model of the outage probability in the environment. In the second step, by using the global model obtained in the first step and rapidly-exploring random trees (RRTs), we propose an algorithm to optimize UAVs' paths. Simulation results show the effectiveness of this two-step approach for UAV networks.
		
	\end{abstract}
	
	
	\IEEEpeerreviewmaketitle
	
	\section{Introduction}
	\IEEEPARstart{U}{NMANNED} aerial vehicles (UAVs) have recently gained significant interest in a broad range of applications. High mobility and flexible deployment are among the features allowing the UAVs to expand their scope of action \cite{RZhang_Survey_Accessing, Azari_Survey, Rahmati_TWC, Behzad_PIMRC_RL}. For effective operation, in many of the applications, the UAVs need to maintain a reliable internet connection during their flights. This connection is mainly provided by ground base stations (GBSs). However, due to location-dependant and time-varying characteristics of the communication channels, and the fact that the cellular networks are designed to serve terrestrial users, service is not available in all parts of the sky \cite{RZhang_cellular_connected_Tcom}. A cost-effective solution to bring this connectivity to the aerial users is to design appropriate trajectories for the UAVs.

	The trajectory optimization problem for cellular-connected UAVs has been extensively studied \cite{RZhang_cellular_connected_Tcom,Guevenc_ICC,Sensing_CommLeter,Behzad_PIMRC_Optimization,Saad_TWC_cellularConnected, RZhang_3Dmap_Globecom}. In \cite{RZhang_cellular_connected_Tcom}, the authors studied the trajectory design problem for a cellular-connected UAV, and based on graph theory and convex optimization, they proposed an algorithm to minimize the mission time of the UAV. This minimization is subject to a maximum tolerable outage duration. In \cite{Guevenc_ICC}, by using dynamic programming, the authors proposed a sub-optimal trajectory design algorithm to minimize the flight time of a cellular-connected UAV. In \cite{Sensing_CommLeter}, the authors proposed an algorithm to optimize the UAV trajectory with the goal of maximizing its energy efficiency. In \cite{Behzad_PIMRC_Optimization}, the authors minimized the propulsion-power consumption of a fixed-wing UAV while a certain connectivity constraint on the instantaneous signal to interference plus noise ratio (SINR) is satisfied. In \cite{Saad_TWC_cellularConnected}, the authors studied the interference management problem for the uplink communication of cellular-connected UAVs. The authors in \cite{Saad_TWC_cellularConnected} obtained the trajectory of the UAVs to achieve a tradeoff between maximizing energy efficiency and minimizing the interference caused on the terrestrial cellular network.

	The studies in \cite{RZhang_cellular_connected_Tcom,Guevenc_ICC,Sensing_CommLeter,Behzad_PIMRC_Optimization,Saad_TWC_cellularConnected, RZhang_3Dmap_Globecom} optimize the trajectory of the UAVs subject to a certain connectivity constraint on the received signal. However, due to the channel randomness, there is no guarantee to satisfy these certain constraints in practice. To take the channel randomness into account, we have to consider probabilistic connectivity constraints. These constraints can be defined in terms of the outage probability and give more reliability to the path planning algorithms. In addition to this concern, the proposed algorithms in \cite{RZhang_cellular_connected_Tcom,Guevenc_ICC,Sensing_CommLeter,Behzad_PIMRC_Optimization,Saad_TWC_cellularConnected, RZhang_3Dmap_Globecom} require instantaneous channel state information. By considering a probabilistic connectivity metric, we can design algorithms that work without having this instantaneous information. To achieve this goal, the UAVs need to have a global model of the outage probability in the environment. Since each UAV has its specific task, and depending on the task, it flies over a particular area, it can not build the global model using its own experience. In contrast, the UAVs need to work together to build this global model.

	To address the issues mentioned above, in this paper, we study the radio mapping and path planning problem for a cellular-connected UAV network. Since the onboard energy of the UAVs is limited, the UAVs can not fly for a long time. To prolong the lifetime of the UAVs, we minimize the flight time of each UAV, ensuring that a probabilistic connectivity constraint is satisfied throughout their flights. This problem is non-convex and challenging to solve. We first reformulate the problem into a mathematically more tractable form. Then, to solve the reformulated problem, we propose a two-step algorithm. In the first step, by using Federated Learning (FL) \cite{FL2, Federated_Google}, the UAVs collaboratively build a global model of the outage probability in the environment. In this method, each UAV uses its data to update the global model locally. Accordingly, the UAVs do not need to share their collected information with a centralized node to do the training task, and the training process is executed in a distributed manner. As a result, the training process will be faster. 
	In the second step, we use the resulting global model to design the trajectory of the UAVs. To achieve this goal, we have to ensure that the connectivity constraint is satisfied. We propose a path planning algorithm based on rapidly-exploring random trees. In our trajectory design algorithm, we do not need the instantaneous channel gains, which was considered in previous works \cite{RZhang_cellular_connected_Tcom,Guevenc_ICC,Sensing_CommLeter,Behzad_PIMRC_Optimization,Saad_TWC_cellularConnected, RZhang_3Dmap_Globecom}. Moreover, our algorithm allows the UAVs to update the model based on their newly collected data. As a result, the model is trained based on online status of the network, which can remarkably increase the model's accuracy. The simulation results show the effectiveness of this two-step approach for UAV networks.

The rest of the paper is organized as follows. In section II, we present the system model and problem formulation. In section III, we propose a two-step algorithm for radio mapping and path planning. Section IV presents the simulation results, and section V concludes the paper.

	\section{System Model and Problem Formulation}
	
	We consider a cellular network, including $J$ GBSs and $U$ cellular-connected UAVs. The UAVs have their missions and fly from their initial locations towards their destinations. 
	We use indices $j$ and $u$ to denote the GBSs and UAVs, respectively. The position of the $u$-th UAV at time $t$ is represented by ${\bf{q}}_u (t) = (x_u(t), y_u (t),h)$, where without loss of generality we assume that the altitude of the UAVs is fixed throughout their flights. The initial and final position of the $u$-th UAV are shown by ${\bf{q}}_u^I$ and ${\bf{q}}_u^F$, respectively. The velocity of the $u$-th UAV at time $t$ and its maximum speed are represented by ${\bf{v}}_u(t) = \frac{d {\bf{q}}_u(t)}{dt}$ and $v_{\text{max}}$, respectively. We denote the position of the $j$-th GBS by ${\bf{q}}^G_j$. Moreover, the flight time of the $u$-th UAV is represented by $T_u$. 
	
	The channel gain between the $u$-th UAV and the $j$-th GBS is expressed as  
	\begin{equation}
	\label{channel}
	h_{u,j} (t) = \frac{\rho_{u,j} (t)}{\text{PL}_{u,j} (t)},
	\end{equation}
	where $\rho_{u,j} (t)$ is the small-scale fading term and ${\text{PL}}_{u,j} (t)$ is the average path-loss (PL) between the $u$-th UAV and the $j$-th GBS at time $t$. The average PL depends on the probability of having a line of sight (LoS) link between the UAV and the GBS. Let $\Gamma_{u,j} (t) \triangleq \left( \frac{4\pi f_c \rVert {\bf{q}}_{u} (t) - {\bf{q}}^G_j \rVert}{c} \right )^2$, where $f_c$ is the carrier frequency and $c$ is the light speed. The average PL is given by
	\begin{equation}
	\label{PL}
	{\bf{PL}}_{u,j} (t) = \Gamma_{u,j} (t) \left(\eta_{\text{LoS}} \xi_{u,j} + \eta_{\text{N-LoS}} (1-\xi_{u,j}) \right),
	\end{equation} 
	where $\eta_{\text{LoS}}$ and $\eta_{\text{N-LoS}}$ are additional losses for the LoS and Non-LoS links, respectively, and $\xi_{u,j}$ is the probability of having a LoS link between the $u$-th UAV and the $j$-th GBS. This probability can be expressed as  $\xi_{u,j} = \left(1 + a\text{exp}(-b(\psi_{u,j}(t)-a))\right )^{-1}$, where $\psi_{u,j}(t)$ is the elevation angle between the $u$-th UAV and the $j$-th GBS at time $t$, and $a$ and $b$ are environment-related parameters \cite{Rahmati_infocom}. According to experimental measurements presented in\cite{Comm_Mag}, for moderate altitudes (less than 100 meters), $\xi_{u,j} \approx 1$ . As a result, \eqref{PL} is simplified to
	\begin{equation*}
	{\bf{PL}}_{u,j} (t) = \Gamma_{u,j} (t) \eta_{\text{LoS}}.
	\end{equation*}
	
	The received signal to interference plus noise ratio (SINR) of the $u$-th UAV from the $j$-th GBS at time $t$ is given by 
	\begin{equation}
	\gamma_{u,j} (t) = \frac{p_j(t) h_{u,j} (t)}{\sum_{j' \neq j} p_{j'}(t) h_{u,j'} (t) + \sigma },
	\end{equation}
	where $\sigma$ is the noise power and $p_j (t)$ is the transmit power of the $j$-th GBS at time $t$.

	As discussed earlier, the UAVs need to maintain reliable communication links to the GBSs. This is essential to support the command and data flows between the UAVs and the cellular network. However, the quality of the link is highly affected by the channel randomness. To take the channel's random characteristic into account, we first define the outage probability as the probability that the received SINR of the UAV from each GBS is less than a certain threshold $\gamma_{\text{th}}$, i.e.,
		 \begin{equation}
	\label{Outage}
	\mathbb{P}_{u}^{\text{outage}} (t) = \mathbb{P}\{ \gamma_{u,j} (t) \leq \gamma_{\text{th}}, \forall j \}  = \mathbb{P}\{ \max_{j} \gamma_{u,j} (t) \leq \gamma_{\text{th}}\}.
	\end{equation}
	Using this outage probability, we can define the connection metric. we say the $u$-th UAV is connected to the cellular network at time $t$ if the outage probability is less than a given threshold $P_0$, i.e., 
\begin{equation}
\mathbb{P}_{u}^{\text{outage}} (t) \leq P_{0}.
\end{equation}
 To have a reliable connection to the cellular network, the UAV is not allowed to loose its connection to the cellular network for more than a given time duration $\delta$. In other words, the maximum continuous time interval that the UAVs can be disconnected from the cellular network is $\delta$. To formulate this constraint, first we define function $\tau_u (t)$ as
\begin{equation}
\label{tau}
\tau_u (t) \triangleq \max \big\{t' \in [0,t]: \mathbb{P}_{u}^{\text{outage}} (t') \leq P_{0} \big\}.
\end{equation}
This function gives the last time instance before $t$ that the $u$-th UAV is connected to the cellular network. Using \eqref{tau}, the reliable connection requirement of the $u$-th UAV is given by \cite{Behzad_ICC}
\begin{equation}
\label{connectivity_constraint}
\max_{t \in [0,T_u]} \{t - \tau_u (t) \} \leq \delta.
\end{equation}
It is worth mentioning that the value of $\delta$ is a design parameter and differs for different applications.

\subsection{Problem Formulation}
The goal of each UAV is to minimize its flight time while its constraint for having a reliable connection to the cellular network is satisfied. If ${\bf{q}}_u= \{ {\bf{q}}_u(t), \forall t \in [0,T_u]\}$, the optimization problem of the $u$-th UAV can be expressed as
	\begin{align*}
	\label{Problem_original}
	\min_{{\bf{q}}_u, T_u}  & \hspace{0.5cm} T_u  \numberthis \\ 
	\text{s.t. }  & \text{C1: } \max_{t \in [0,T_u]} \{t - \tau_u (t)\} \leq \delta, \\
	& \text{C2: } \begin{matrix}  \left \rVert \frac{d{\bf{q}}_u (t)}{dt} \right \rVert \leq v_{\text{max}}, \hspace{0.5cm} \forall t \in [0,T_u], \end{matrix} \\
	& \text{C3: } \begin{matrix}  {\bf{q}}_u(0)={\bf{q}}_u^{\text{I}}, \text{ and } {\bf{q}}_u(T_u)={\bf{q}}_u^{\text{F}}.\end{matrix}
	\end{align*}
    In \eqref{Problem_original}, constraint C1 shows the reliable connection requirement of the UAV. Constraint C2 states that the UAV's velocity is limited to its maximum speed, and constraint C3 represents the initial and final location of the UAV. Problem \eqref{Problem_original} is non-convex due to C1. Moreover, the outage probability used in $\tau_u (t)$ depends on the network's topology. Even if we consider a simple topology for the network, the UAVs do not have access to the outage probability in the environment. As a result, traditional optimization techniques can not be used to solve \eqref{Problem_original}. In addition to these concerns, the UAV's flight time, $T_u$, is among the optimization variables. To solve \eqref{Problem_original}, we have to find $T_u$ and the value of ${\bf{q}}_u(t)$ for all $t \in [0, T_u]$, which is a challenging task. In what follows, we reformulate the problem into a more tractable form. With this reformulation, we do not need to solve the continuous-time problem. Instead, we can find the solution of an equivalent discrete-time problem.

	\subsection{Problem Reformulation}
	The goal of each UAV is to solve its corresponding optimization problem. For the sake of brevity, we omit index $u$ from the problem and continue our discussion for the general case. To reformulate problem \eqref{Problem_original},  we use the fact that any feasible solution must satisfy constraint C1. Hence, instead of solving the problem for all time instances, it is sufficient to consider the problem for a sequence of discrete time instances $t_1$, $t_2$, $\ldots$, $t_N$, where $| t_n - t_{n-1}| \leq \delta$, $n=1, \ldots, N$, and make sure that the UAV is connected to the cellular network at these time instances. In other words, if  
		\begin{equation}
	\mathbb{P}_{\text{outage}}(t_n)  \leq P_0, \hspace{0.5cm} n=1, \ldots, N,
	\end{equation}
	the maximum continuous time that the UAV is in outage will be limited to $\delta$. Therefore, constraint C1 will be satisfied. To reformulate C2, we have $\left \rVert \frac{d{\bf{q}} (t)}{dt} \right \rVert = 
	\frac{\rVert {\bf{q}} (t_n) - {\bf{q}} (t_{n-1}) \rVert}{|t_n - t_{n-1}|}$, Since $| t_n - t_{n-1}| \leq \delta$, the equivalent form of C2 is given by
	\begin{equation}
	\label{C2:equivalent}
	 \rVert {\bf{q}} (t_n) - {\bf{q}} (t_{n-1}) \rVert \leq   \delta v_{\text{max}}, \hspace{0.5cm} n=1, \ldots, N.
	\end{equation}
	It can be shown that to minimize the flight time, the UAVs fly with their maximum speed, $v_{\text{max}}$. Considering this fact and using a similar approach to what has been shown for C2, we can write the objective of \eqref{Problem_original} as
	\begin{equation}
	T = \frac{1}{v_{\text{max}}}\sum_{n=1}^{N} \left\rVert {\bf{q}} (t_n) - {\bf{q}} (t_{n-1}) \right \rVert.
	\end{equation}
	Let ${\bf{q}}[n] \triangleq {\bf{q}}(t_n)$, $n=0, 1, \ldots, N$. Problem \eqref{Problem_original} is equivalent to the following discrete-time optimization problem  
	\begin{align*}
	\label{Problem_reformulated}
	\min_{\{{\bf{q}}[n]\}_{n=0}^{N}, N}  & \hspace{0.5cm} \sum_{n=1}^{N} \rVert {\bf{q}}[n] - {\bf{q}}[n-1] \rVert  \numberthis \\ 
	\text{s.t. }  & \tilde{\text{C}}\text{1: } \mathbb{P}^{\text{outage}}({\bf{q}}[n])  \leq P_0, \forall n, \\
	& \tilde{\text{C}}\text{2: } \begin{matrix}  \rVert {\bf{q}}[n] - {\bf{q}}[n-1] \rVert \leq \delta v_{\text{max}}, \end{matrix} \\
	& \tilde{\text{C}}\text{3: } \begin{matrix}  {\bf{q}}[0]={\bf{q}}^{\text{I}}, \text{ and } {\bf{q}}[N]={\bf{q}}^{\text{F}}.\end{matrix}
	\end{align*}
	\section{Two-step Algorithm for Radio Mapping and Path Planning}
	To find the solution of the reformulated problem, we still need to know the outage probability. However, this information is not availbale to the UAVs. Hence, we need to obtain this information first. In what follows we propose a two-step algorithm to solve \eqref{Problem_reformulated}. In this approach, in the first step, we estimate the outage probability. To achieve this goal, we use Federated Learning (FL) which allows the UAVs to collaborate to build a global model of the outage probability based on their locally collected data. Moreover, the UAVs do not need to share their collected data with a central node in this approach. In the second step, we use the derived model to find the solution of the problem ensuring that $\tilde{\text{C}}\text{1}$ is satisfied.

		\subsection{Radio Mapping based on Federated Learning}
		
		\begin{figure}[t]
		\centering
		\includegraphics[trim={0cm 0cm 0cm 0cm}, width=3.2in,keepaspectratio]{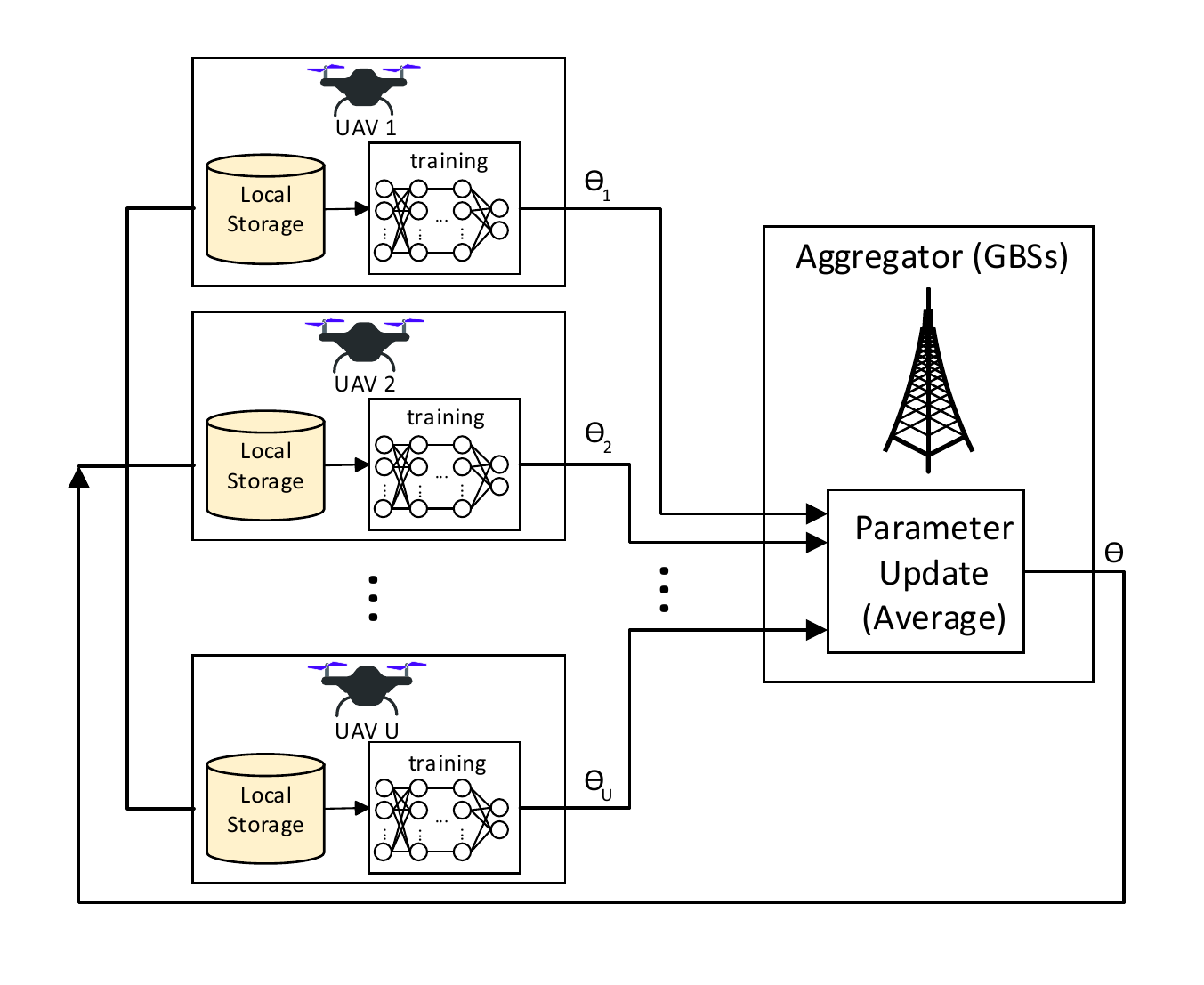}
		\vspace{-0.3cm}
		\caption{Federated Learning to estimate the outage probability.}
		\label{fig:1}
	\end{figure}

	To estimate $\mathbb{P}^{\text{outage}}$(.), as depicted in Fig. \eqref{fig:1}, we consider a scenario where all UAVs can collaborate to build a global model of the outage probability in the environment. Let $\mathcal{D}_u$ denote the data set of the $u$-th UAV and let  $\mathcal{D} = \cup_{u=1}^{U} \mathcal{D}_u$. We denote $|\mathcal{D}_u|=m_u$ and $|\mathcal{D}|=M=\sum_{u=1}^{U} m_u$, where $|A|$ is the cardinality of set $A$. The data set of the $u$-th UAV includes all pairs of $({\bf{q}}_u (t), l_{u}(t))$, where ${\bf{q}}_u (t)=(x_u(t),y_u(t))$ is the 2D coordinate of a visited location by the UAV and 
	\begin{equation}
	l_u (t) = \begin{cases}
	1 & \text{if } \max_{j} \gamma_{u,j} (t) \geq \gamma_{\text{th}},\\
	0 & \text{otherwise},
	\end{cases}
	\end{equation} 
	is the label of this data point. To find a global model of the outage probability, we consider a neural network with parameter $\theta$. The output of this neural network for input $\bf{q} \in \mathbb{R}^2$ is denoted by $P(\theta,{\bf{q}})$ which gives the outage probability at location ${\bf{q}}$. To find an appropriate model for the outage probability, we have to solve the following problem
\begin{equation}
	\label{FL:problem:main}
	\min_{\theta}  f(\theta )  = \frac{1}{M}\sum_{m=1}^{M} f_m (\theta),
	\end{equation}
	where $f_m$ is the loss function corresponding to the $m$-th data point. We consider cross-entropy for the loss function as
	\begin{align}
	\label{CrossEntropy}
	f_m (\theta) \triangleq & f(\theta; {\bf{q}}^m,l^m) \\
	= &  - l^m \log (P(\theta,{\bf{q}}^m)) - (1-l_m)\log (1-P(\theta,{\bf{q}}^m)), \nonumber
	\end{align}
	where $\left({\bf{q}}^m, l^m\right)$ is the $m$-th data point. We can write the objective of \eqref{FL:problem:main} as
	\begin{equation}
	f(\theta) = \frac{1}{M} \sum_{u=1}^{U}   \sum_{({\bf{q}}^m,l^m) \in \mathcal{D}_u} f(\theta; {\bf{q}}^m,l^m).
	\end{equation}
	If we define $F_u (\theta) = \frac{1}{m_u} \displaystyle \sum_{({\bf{q}}^m,l^m) \in \mathcal{D}_u} f(\theta; {\bf{q}}^m,l^m)$, then we have
	\begin{equation}
	\label{FL:reformulated_objective}
	f(\theta) = \sum_{u=1}^{U} \frac{m_u}{M} F_u (\theta).
	\end{equation}
		\begin{algorithm}[t]
		\footnotesize
		\label{FL}
		\caption{FL-based radio mapping to estimate outage probability.}
		The GBSs initialize the global model parameter, $\theta$,
		randomly.\\
		set $t=1$ and $\theta^{(t)}=\theta$.\\
		\For{$t=1$ to $T_{FL}$ (max round)}{ 
			The GBSs send the global model parameter, $\theta^{(t)}$, to the UAVs\\
			\For {each UAV $u$} { 
				$\theta_u = \theta^{(t)} $\\
				Update training data set $\mathcal{D}_u$ using newly collected data\\
				\For{$epoch = 1$ to $H$}{
					Update its local parameter $\theta_u$ over its data set $\mathcal{D}_u$ as\\
					\[\theta_u = \theta_u - \eta \nabla F_u (\theta_u)\] 
				}
				Each UAV sends its local parameter to the cellular network.\\}
			The GBSs collect all local parameters and updates the global model as
			\[ \theta^{(t+1)} = \displaystyle \sum_{u=1}^{U} \frac{m_u}{M}\theta_u  \]
		}
	\end{algorithm}
    To minimize \eqref{FL:reformulated_objective}, instead of using a central approach which requires access to data of all UAVs, we use a distributed approach based on FL \cite{FL2}. In this algorithm, we assume that the global model parameter, $\theta$, is available to all GBSs. The GBSs send this parameter to the UAVs and the UAVs update their local model parameter as $\theta_u = \theta, \forall u$. The UAVs fly over the area and based on their received signals from the cellular network, they form their data sets $\mathcal{D}_u, \forall u$. Using the collected data, each UAV performs $H$ steps of the stochastic gradient descent (SGD) on its local parameter, $\theta_u$. In other words, in each step, the $u$-th UAV updates its local parameter as
	\begin{equation}
	\theta_u = \theta_u - \eta \nabla F_u (\theta_u),
	\end{equation} 
	where $\eta$ is the step size and $\nabla$ is the gradient operator. The updated parameters, $\theta_u, \forall u$, are sent back from the UAVs to the GBSs. The GBSs act as aggregators and share this infromation in their network. By averaging the received local parameters, the GBSs evaluate a new parameter as
	\begin{equation}
	\theta= \displaystyle \sum_{u=1}^{U} \frac{m_u}{M}\theta_u .
	\end{equation}
	The global model is updated using this new $\theta$. This updated parameter is again sent to all UAVs to perform their local updates with their new data. This procedure is repeated until the neural network is trained. Algorithm 1 presents the FL-based radio mapping for outage probability estimation. 
	\begin{algorithm}[t]
		\footnotesize
		\label{RRT_star}
		\caption{RRT$^*$-based path planning for the UAVs }
		$V = \{{\bf{q}}^I\}$, $E=\{\}$, $\mathcal{T}= (V,E)$\\
		\For{$n =1$ \text{ to } $N$ }{
			Randomly sample a point from the space ${\bf{q}}_{\text{rand}}$\\
			Find the nearest vertex of the graph to ${\bf{q}}_{\text{rand}}$, i.e.,
			\begin{equation*}
			{\bf{q}}_{\text{nearest}} = \displaystyle \argmin_{ {\bf{q}}\in V} \rVert {\bf{q}}_{\text{rand}} - {\bf{q}} \rVert
			\end{equation*}  
			Find ${\bf{q}}_{\text{new}} = \argmin_{\rVert {\bf{q}} - {\bf{q}}_{\text{nearest}} \rVert \leq v_{\text{max}} \delta} \rVert {\bf{q}}_{\text{rand}} - {\bf{q}} \rVert $\\
			\If{$P(\theta^*, {\bf{q}}_{\text{new}}) \leq P_0$}{ 
				$Q_{\text{near}} = \{{\bf{q}} \in V : \rVert {\bf{q}} - {\bf{q}}_{\text{new}} \leq v_{\text{max}} \delta\}$\\
				$V = V \cup \{{\bf{q}}_{\text{new}}\}$\\
				${\bf{q}}_{\text{min}} = {\bf{q}}_{\text{nearest}}$\\
				$c_{\text{min}} = c({\bf{q}}_{\text{nearest}}) + \rVert {\bf{q}}_{\text{nearest}} - {\bf{q}}_{\text{new}} \rVert$\\
				\For{all ${\bf{q}}_{\text{near}} \in Q_{\text{near}}$}{
					\If{$P(\theta^*,{\bf{q}}_{\text{near}}) \leq P_0$ and $c({\bf{q}}_{\text{near}}) + \rVert {\bf{q}}_{\text{near}} - {\bf{q}}_{\text{new}} \rVert < c_{\text{min}}$}{
						${\bf{q}}_{\text{min}} = {\bf{q}}_{\text{near}}$\\
						$c_{\text{min}} = c({\bf{q}}_{\text{near}}) + \rVert {\bf{q}}_{\text{near}}-{\bf{q}}_{\text{new}}\rVert$\\
					}		
				}
				$E = E \cup \{({\bf{q}}_{\text{min}},{\bf{q}}_{\text{new}} )\}$\\		
				\For{all ${\bf{q}}_{\text{near}} \in Q_{\text{near}}$}{
					\If{$P(\theta^*,{\bf{q}}_{\text{near}}) \leq P_0$ and $c({\bf{q}}_{\text{near}}) + \rVert {\bf{q}}_{\text{near}} - {\bf{q}}_{\text{new}} \rVert < c({\bf{q}}_{\text{near}})$}{
						Find the parent of ${\bf{q}}_{\text{near}}$, i.e.,\\
						${\bf{q}}_{\text{p}} = \{{\bf{q}} : ({\bf{q}},{\bf{q}}_{\text{near}}) \in E\}$	\\
						$E = \{E \setminus \{({\bf{q}}_{\text{p}},{\bf{q}}_{\text{near}})\} \} \cup \{({\bf{q}}_{\text{new}}, {\bf{q}}_{\text{near}})\}$
					}
				}	
			}
		} 
	\end{algorithm}
	The advantage of this approach is that the network parameter can be locally updated. This does not require high computation resources. Moreover, the UAVs do not need to share their data sets with the cellular network. In addition to these benefits, the UAVs might fly over differnt areas of the network. Hence, they will not have good knowledge of the other parts of the network. This FL-based algorithm allows the UAVs to collaborate to build a global model based on their limited local experiences.

	\subsection{Path Planning based on RRT$^{*}$}
	After the first step, we have a model of the outage probability in the environment. Let $\theta^*$ denote the parameter of the trained global model. In this step, we have to solve the following optimization problem
	\begin{align*}
	\label{RRT:Problem}
	\min_{\{{\bf{q}}[n]\}_{n=0}^{N}, N}  & \hspace{0.5cm} \sum_{n=1}^{N} \rVert {\bf{q}}[n] - {\bf{q}}[n-1] \rVert  \numberthis \\ 
	\text{s.t. }  & \tilde{\text{C}}\text{1: } P(\theta^*, {\bf{q}}[n])  \leq P_0, \hspace{0.5cm} n=0,1, \ldots, N, \\
	& \tilde{\text{C}}\text{2-}\tilde{\text{C}}\text{3, } 
	\end{align*}
	where $P(\theta^*,{\bf{q}})$ is the output of the trained neural network for position ${\bf{q}}$. Problem \eqref{RRT:Problem} is still challenging to solve. The reason is that the feasible region corresponding to constraint $\tilde{\text{C}}$1 is not necessarily convex. Moreover, $N$ which is the number of discrete time steps is a variable of the problem. To overcome these difficulties, we use rapidly-exploring random trees (RRTs) \cite{Karaman}. RRTs are designed to search non-convex spaces. In what follows, we describe our algorithm which works based on a modified version of RRTs called RRT$^{*}$ \cite{Karaman}. RRT$^{*}$ is an efficient algorithm which can find the shortest path between a pair of initial and final locations in a continuous space.
	 
		\begin{figure}
		\begin{subfigure}{.23\textwidth}
			\centering
			\includegraphics[trim={2cm 0 2.cm 0.8cm},clip, width=1\linewidth]{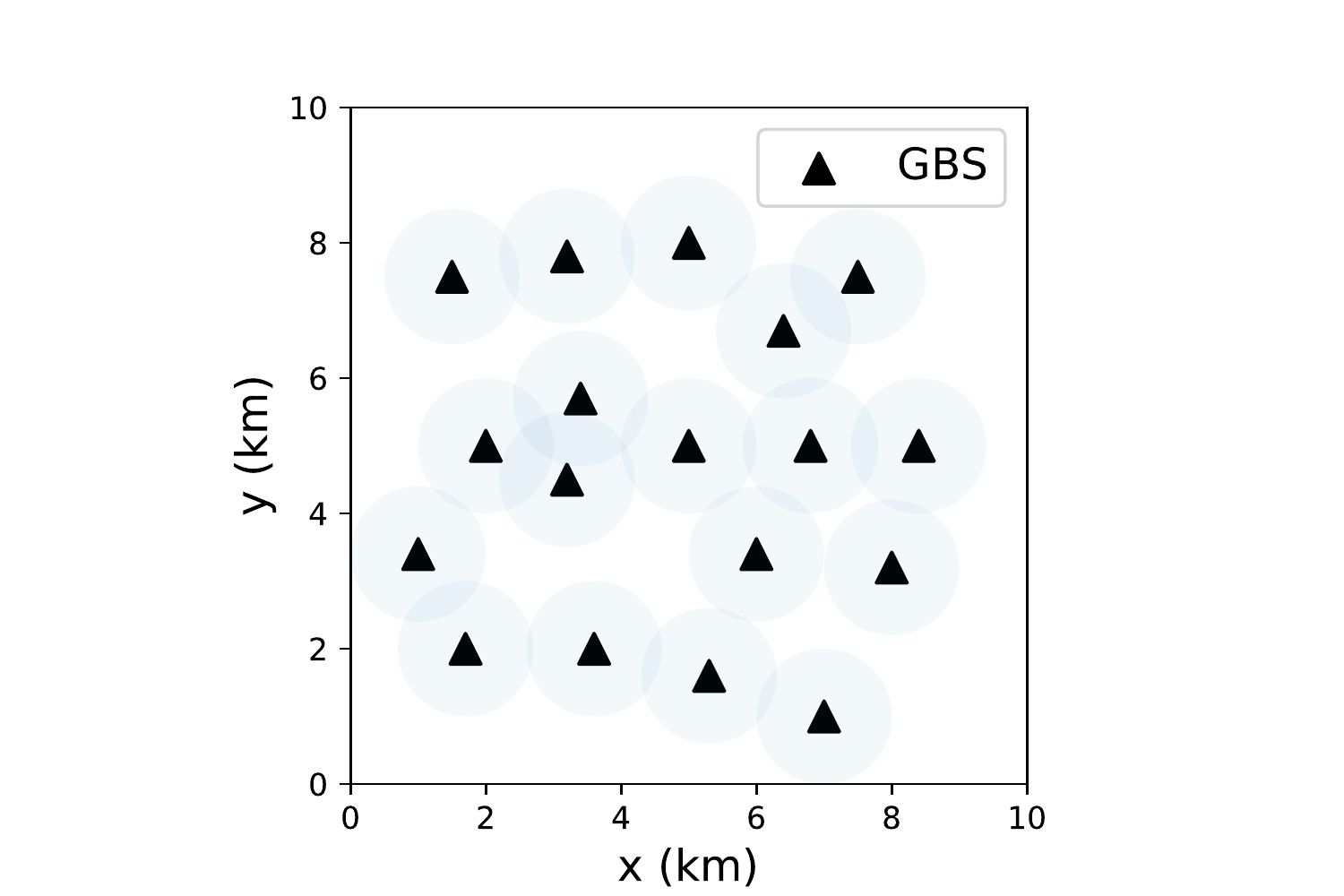}  
			\caption{}
			\label{fig2:1}
		\end{subfigure}
		\begin{subfigure}{.23\textwidth}
			\centering
			\includegraphics[trim={1.8cm 0 0.5cm 0},clip, width=1.1\linewidth]{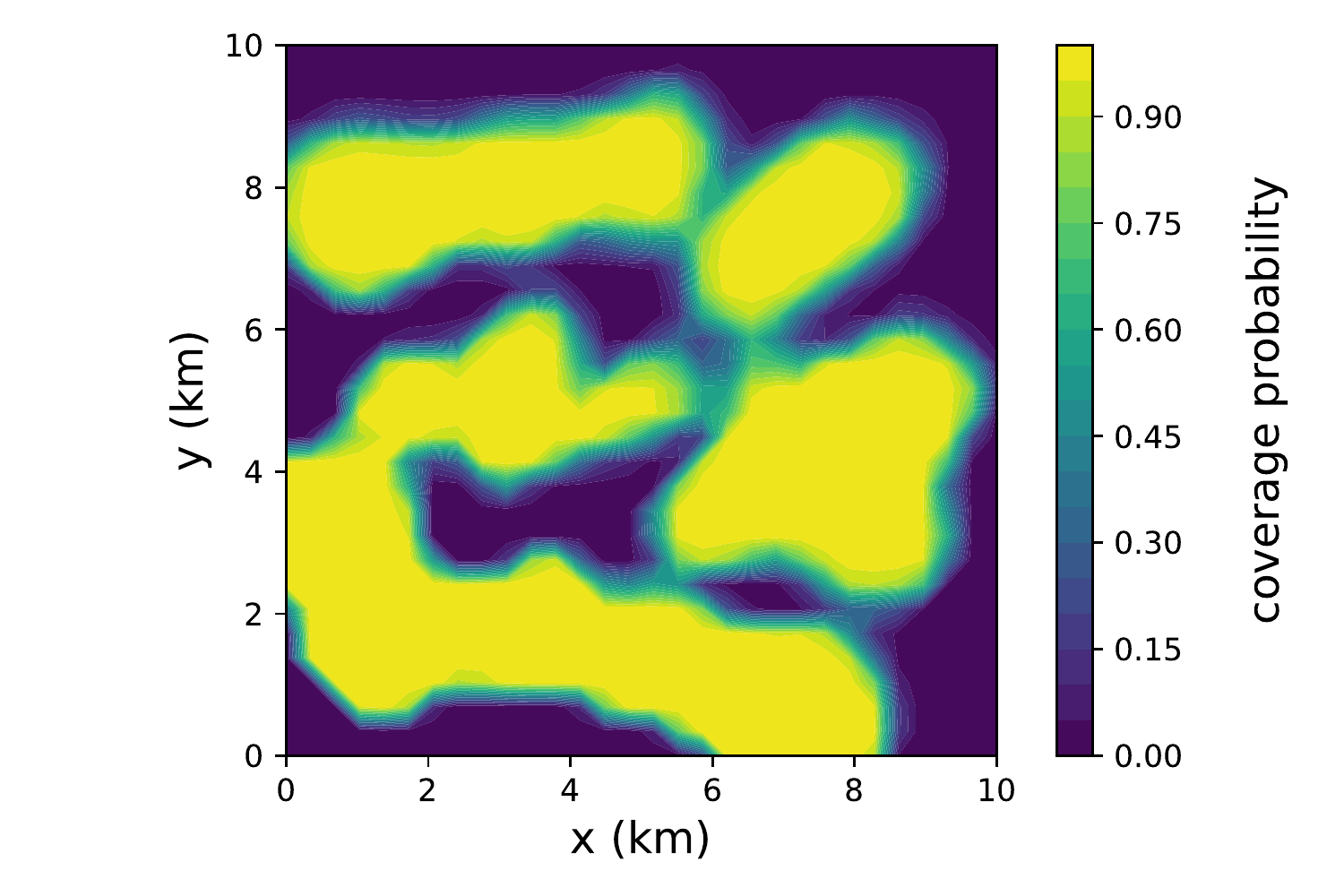} 
			\caption{}
			\label{fig2:2}
		\end{subfigure}
	\vspace{-0.1cm}
		\caption{ (a) Location of the GBSs (b) Trained outage probability}
		\label{fig2}
	\end{figure}

	We use RRT$^*$ to find the path that minimizes the UAV's flight time while ensuring that it satisfies the requirement for having a reliable connection to the cellular network. In our algorithm, we use a tree $\mathcal{T} = (V,E)$ to represent the path between the initial point to any feasible coordinate ${\bf{q}}$. The set of vertices, $V$, is the set of all coordinates ${\bf{q}}$ in the space that has been explored by the algorithm. The root of the tree is the initial point ${\bf{q}}^{I}$. Coordinate ${\bf{q}}_p$ is called parent of ${\bf{q}}_c$ if the UAV arrives in ${\bf{q}}_c$ from ${\bf{q}}_p$ and it does not visit any other vertex in between. According to this notation, $E$ is the set of all edges $({\bf{q}}_p,{\bf{q}}_c)$ which connects a visited coordinate to its parent. Moreover, in our algorithm, we need to define a cost function $c(.)$ for each visited coordinate. If the UAV passes the sequence of edges $\{({\bf{q}}_0, {\bf{q}}_1), ({\bf{q}}_1,{\bf{q}}_2), \ldots, ({\bf{q}}_{n-1},{\bf{q}}_n)\}$ to reach ${\bf{q}}_n$, where ${\bf{q}}_0 = {\bf{q}}^I$, then the cost of coordinate ${\bf{q}}_n$ is defined as 
	\begin{equation}
	c({\bf{q}}_n) = \sum_{n'=1}^{n} \rVert {\bf{q}}_{n'} - {\bf{q}}_ {n'-1} \rVert. 
	\end{equation}
		
We start our path planning algorithm from the root, ${\bf{q}}^I$, and iteratively add new vertices to the tree and update its structure including the edges and costs. In each iteration, to add a new vertex to the tree, we sample a random coordinate from the space. This random coordinate is denoted by ${\bf{q}}_{\text{rand}}$. Then, the closest vertex of $\mathcal{T}$ to ${\bf{q}}_{\text{rand}}$ is found and is shown as ${\bf{q}}_{\text{nearest}}$. To satisfy constraint C2, we have to ensure that the distance between the new sampled coordinate and ${\bf{q}}_{\text{nearest}}$ is less than $\delta v_{\text{max}}$, i.e., $\rVert {\bf{q}}_{\text{rand}} - {\bf{q}}_{\text{nearest}} \rVert \leq \delta v_{\text{max}}$. To meet this requirement, in case that this inequality is violated, we replace ${\bf{q}}_{\text{rand}}$ with a closer point denoted as ${\bf{q}}_{\text{new}}$, where
	\begin{equation*}
{\bf{q}}_{\text{new}} = \argmin_{\rVert {\bf{q}} - {\bf{q}}_{\text{nearest}} \rVert \leq \delta v_{\text{max}}} \rVert {\bf{q}}_{\text{rand}} - {\bf{q}} \rVert.
\end{equation*}
After satisfying C2, we have to make sure that C1 is also satisfied. If $P(\theta^*, {\bf{q}}_{\text{new}}) \leq P_0$, then this new coordinate ${\bf{q}}_{\text{new}} $ is out of outage and hence, it can be added to the tree. Otherwise, we have to repeat this iteration from the begining to find a coordinate satisfying both C1 and C2. Assuming $P(\theta^*, {\bf{q}}_{\text{new}}) \leq P_0$, we add this point to the tree as a new vertex. To set the cost of this new vertex, we find all vertices of $\mathcal{T}$ whose distance to ${\bf{q}}_{\text{new}}$ is less than $\delta v_{\text{max}}$. From these vertices, we choose the one minimizing 
\begin{equation}
\label{min:cost}
{\bf{q}}_{\text{min}}= \argmin_{{\bf{q}} : {\bf{q}} \in V {\text{ and }} \rVert {\bf{q}} - {\bf{q}}_{\text{new}} \rVert \leq \delta v_{\text{max}}} c({\bf{q}}) + \rVert {\bf{q}} - {\bf{q}}_{\text{new}} \rVert.
\end{equation}
After finding ${\bf{q}}_{\text{min}}$, we add edge $({\bf{q}}_{\text{min}}, {\bf{q}}_{\text{neq}})$ to $E$ and rewire the tree to update the parents of the vertices located in a distance less than $\delta v_{\text{max}}$ to ${\bf{q}}_{\text{new}}$. Algorithm 2 presents the path planning procedure for the UAVs.

\begin{figure}[t]
	\centering
	\includegraphics[trim={0cm 0cm 0cm 0cm}, width=2.4in,keepaspectratio]{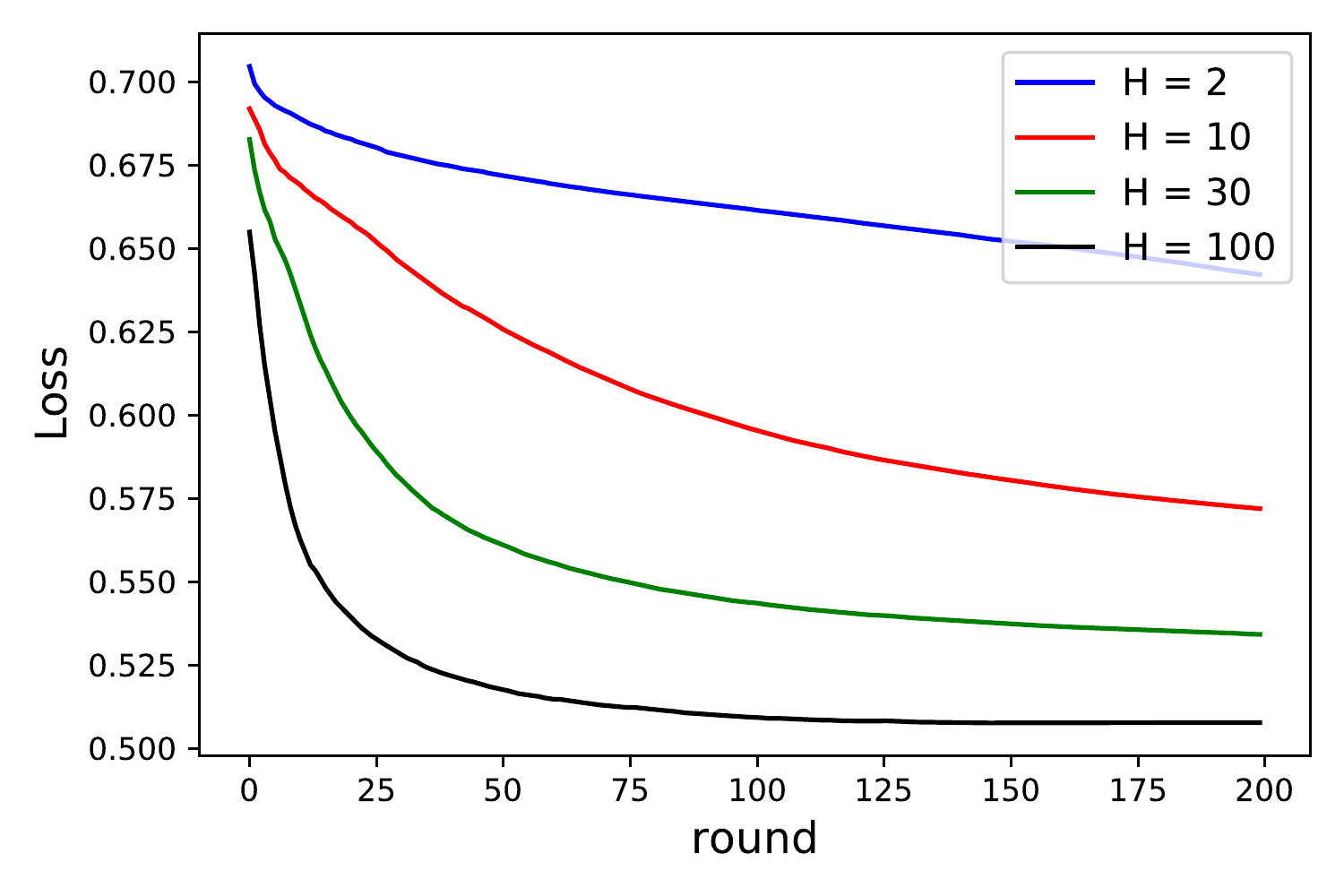}
	\caption{Loss function for different values of $H$.}
	\vspace{-0.3cm}
	\label{fig:3}
\end{figure}	

\begin{figure}[t]
	\centering
	\includegraphics[trim={0cm 0cm 0cm 0cm}, width=2.4in,keepaspectratio]{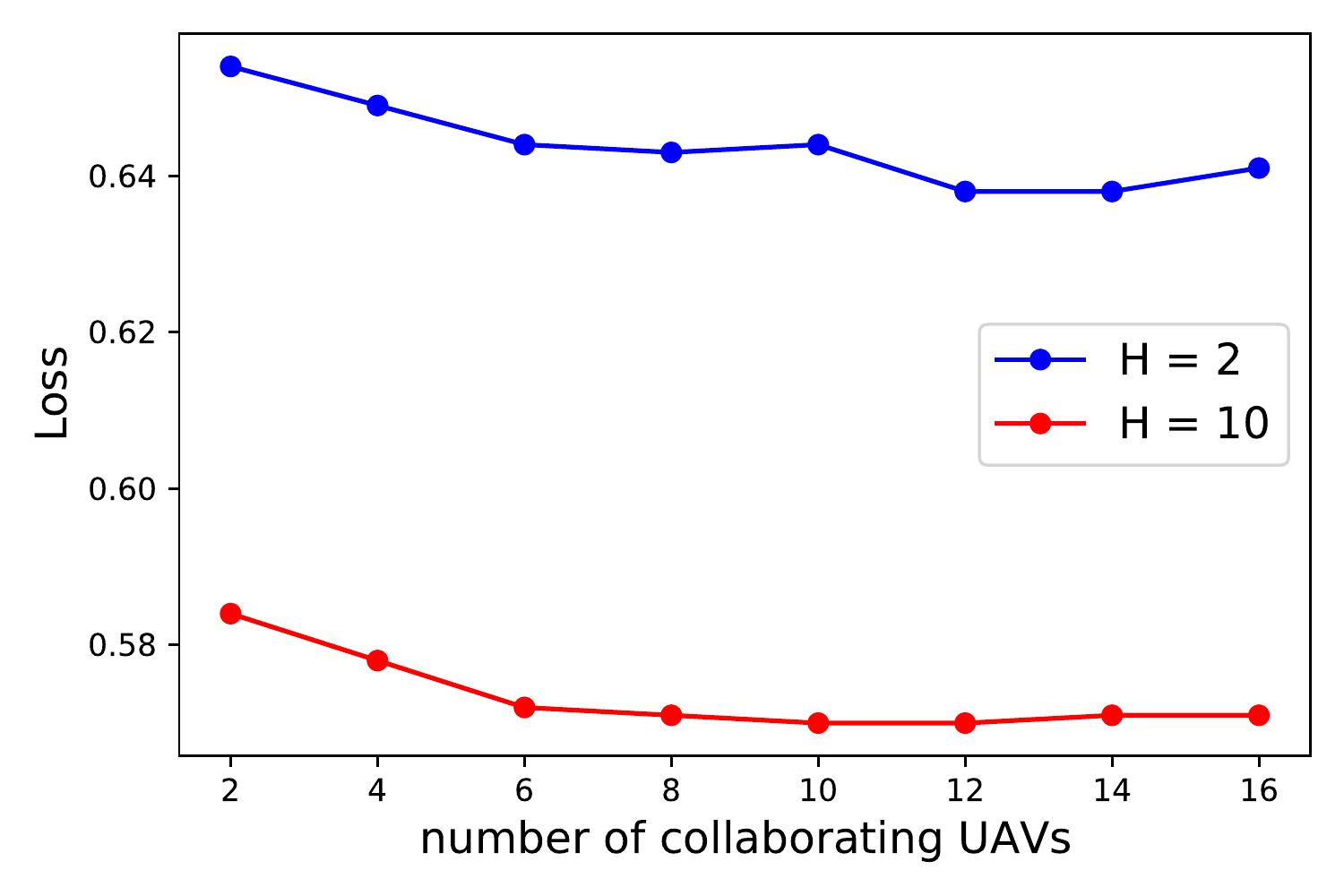}
	\caption{Loss function for different number of collaborating UAVs.}
	\label{fig:4}
\end{figure}

	\begin{figure*}[t]
	\begin{subfigure}{.31\textwidth}
		\centering
		\includegraphics[trim={2cm 0 0.5cm 0cm},clip, width=1.05\linewidth]{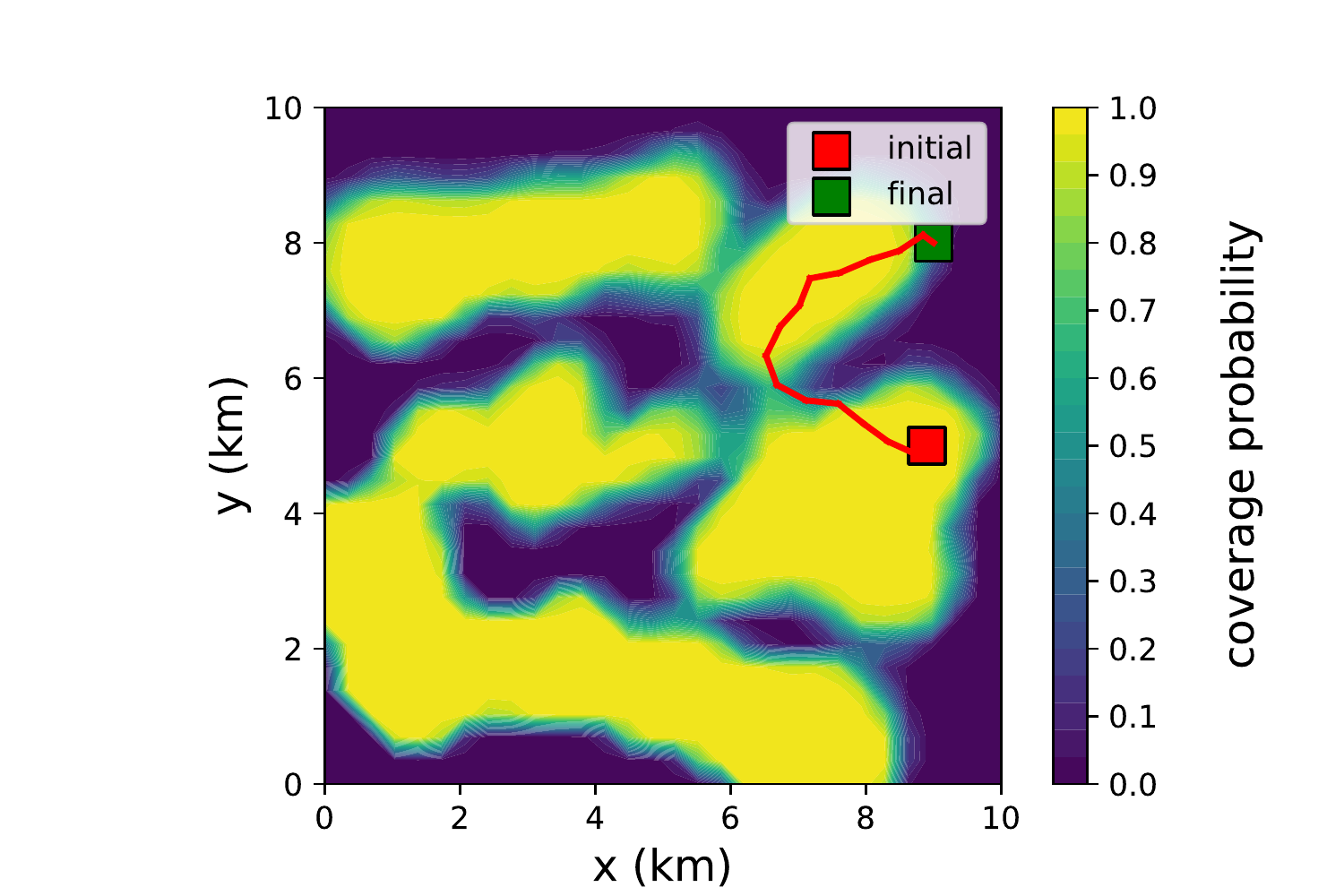}  
		\caption{}
		\label{fig5:1}
	\end{subfigure}
\hfill
	\begin{subfigure}{.31\textwidth}
		\centering
		\includegraphics[trim={2cm 0 0.5cm 0},clip, width=1.05\linewidth]{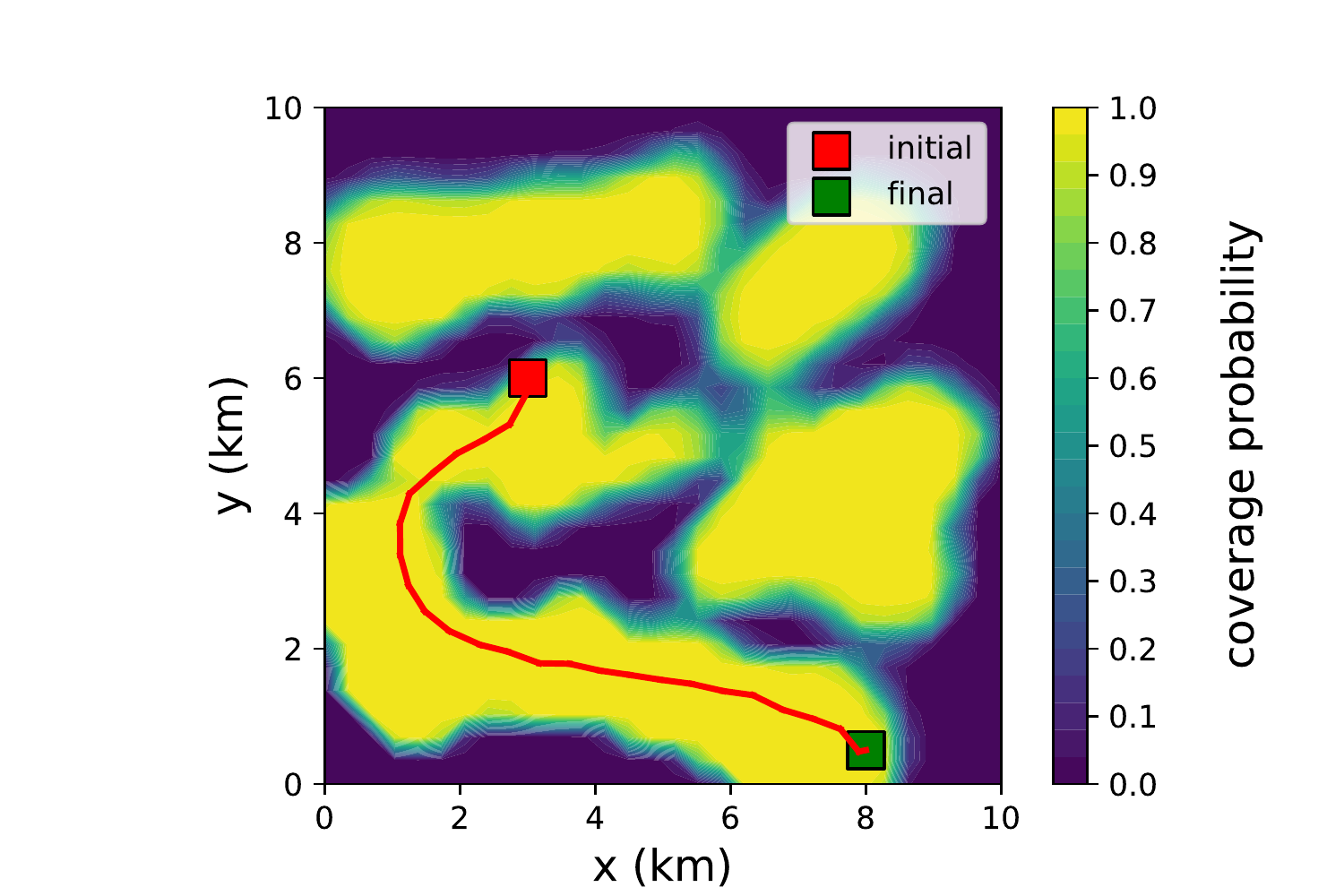} 
		\caption{}
		\label{fig5:2}
	\end{subfigure}
\hfill
	\begin{subfigure}{.31\textwidth}
		\centering
		\includegraphics[trim={2cm 0 0.75cm 0},clip, width=1.02\linewidth]{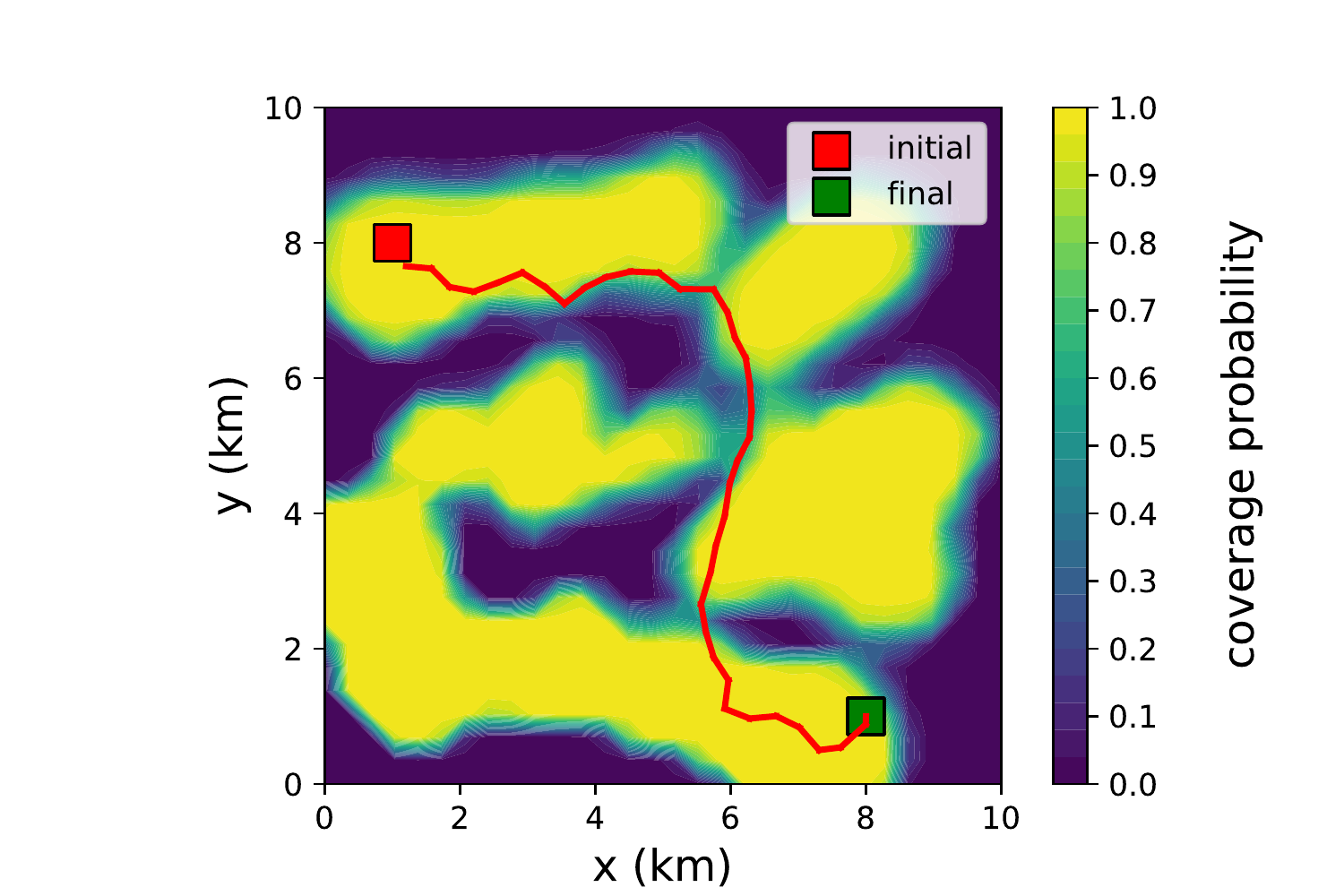} 
		\caption{}
		\label{fig5:3}
	\end{subfigure}
	\caption{ Path of the UAV for different pairs of initial and final locations when (a) $P_0=0.4$  (b) $P_0=0.05$ (c) $P_0=0.30$. }
	\label{fig:5}
	\vspace{-0.5cm}
\end{figure*} 
	\section{Simulation Results}
	In this section, we present the simulation results to show the performance of the proposed algorithm. We consider a $10$km $\times$ $10$km area as shown in Fig. \eqref{fig2:1}. The altitude of the UAVs is $100$m. The transmit power of each GBS is $200$mW. For the fading term, we consider a Rayleigh random variable with parameter $1$. Other simulation parameters are: $v_{\text{max}} = 5\frac{\text{m}}{\text{s}}$, $\sigma = -140$dB, $(a,b) = (5,0.5)$, $f_c=2$GHz, $\eta_{\text{LoS}}=1$,  and $\eta_{\text{N-LoS}}=20$. Moreover, we consider $\gamma_{\text{th}}=0.65$ and $\delta = 5$s.
	
	To estimate the outage probability, we use a fully-connected neural network. The neural network has three hidden layers. Each hidden layer consists of $256$ neurons. The last layer of the neural network (output layer) has two outputs. We use rectified linear unit (ReLU) activation functions for the hidden layers. For the output layer, we use a softmax activation function. The batch size of each UAV is $500$. 
	
	Fig. \eqref{fig2:2} shows the resulting coverage (1-outage) probability in the environment. The number of collaborating UAVs is $10$, the number of training rounds, $T_{FL}$, is $200$, and the number of SGD updates in each round is $10$. As can be seen, in locations close to the GBSs, we expect to have stronger signals, which lead to a lower outage probability. However, as the distance from the GBSs increases, due to the PL, the outage probability highly increases. It is worth mentioning that to obtain this model, the UAVs do not know the true location of the GBSs. The only information they have is their received signals in different locations. Fig. \eqref{fig2:2} also shows that connectivity is not available in all parts of the sky. So, it is essential to design proper paths for the UAVs.

	Fig. \eqref{fig:3} shows the average loss for different numbers of training steps per round ($H$). We observe that as the value of $H$ increases, the UAVs have more time to update their models using the same training data. Therefore, the loss will decrease. 
	
	Fig. \eqref{fig:4} presents the average loss for different numbers of collaborating UAVs. We observe that as the number of collaborating UAVs increases, the loss function decreases. However, this effect is negligible compared to the impact of $H$ on the loss function. In fact, by using more UAVs, we get data from different areas of the environment, depending on the flight paths of the UAVs. To use this data, the UAVs need to have enough updates on their model. If the value of $H$ is small, increasing the number of collaborating UAVs does not significantly increase the global model's accuracy.

    Fig. \eqref{fig:5} presents the trajectory of a UAV for three different pairs of initial and final locations. The values of $P_0$ in (a), (b), and (c) are $0.4$, $0.05$ and $0.30$, respectively. To obtain these paths, Algorithm 2 considers coordinates $\bf{q}$ with $P(\theta^*,{\bf{q}}) \leq P_0$. We observe that to satisfy the connectivity constraint, the UAV needs to take longer paths which increases its flight time.

	\section{conslusion}
    In this paper, we studied the radio mapping and path planning problem for a UAV network. We minimized the UAVs' flight time, ensuring that the UAVs satisfy a probabilistic connectivity constraint during their flights. To solve the problem, we proposed a two-step approach. In the first step, using FL, the UAVs build a global model of the outage probability in the environment based on their collected data. In the next step, using this learned model and rapidly-exploring random trees, we develop a path-planning algorithm that satisfies the cellular-connectivity requirements.	
	
	\bibliographystyle{IEEEtran}
	\bibliography{Citations}
	
\end{document}